\documentstyle[11pt,newpasp,twoside,epsfig]{article}
\def\edcomment#1{\iffalse\marginpar{\raggedright\sl#1\/}\else\relax\fi}
\reversemarginpar

\begin{document}
\title{Implications of the Optical Observations of Isolated Neutron Stars}
 \author{Andy  Shearer, Aaron Golden }
\affil{The National University of Ireland, Galway, Newcastle Road, Galway, Ireland}
\author{Gregory Beskin}
\affil{Special Astrophysical Observatory, Russia}

\begin{abstract}

Observations of the 5 confirmed optical pulsars indicate that the peak
emission scales according to the outer field strength. We show that
this gives gives further confirmation that a simple phenomenological
models such as Pacini and Salvati (1987) still have
validity. Furthermore we show that the Crab pulsar exhibits unpulsed
emission which further complicate any studies of the thermal emission
from Isolated Neutron Stars.

 \end{abstract}

\section{Introduction}

Since the first optical observations of the Crab pulsar in the late
1960s (Cooke et al 1969) only 4 more pulsars have been seen to pulsate
optically (Middleditch and Pennypacker (1985); Wallace et al (1977); Shearer et al (1997); Shearer et al (1998)). Four of these five
pulsars are probably too distant to have any detectable thermal
emission. For the fifth (and faintest) pulsar, PSR 0633+17, emission
has been shown to be non-thermal (Martin et al 1998). Many suggestions
have been made concerning the optical emission process for these young
and middle-aged pulsars. However
the most successful phenomenological model (both in terms of its
simplicity and longevity) has the proposed by Pacini (1972) and in
modified form by Pacini and Salvati (1983, 1987). In general they
proposed that the high energy emission comes from electrons radiating
via synchrotron processes in the outer regions of the magnetosphere.
               
In recent years a number of groups of carried out detailed simulations
of the high-energy processes. These models divide into two groups -
between emission low in the magnetosphere (polar cap models) and those
with the acceleration nearer to the light cylinder (outer-gap
models). Both models have problems explaining the observed features of
the limited selection of high energy emitters. Both models suffer from
arbitrary assumptions in terms of the sustainability of the outer-gap
and the orientation of the pulsar's magnetic field to both the
observers line of sight and the rotation axis. Furthermore observational
evidence, see for example Eikenberry \& Fazio (1997), severely limits
the applicability of the outer-gap to the emission from the
Crab. However they have their successes - the total polar-cap
emission can be understood in terms of the Goldreich and Julian
current from in or around the cap; the Crab polarisation sweep is
accurately produced by an outer-gap variant Romani et al (1995).
               
It is the failure of the detailed models to explain the high energy
emission which has prompted this work. We have taken a
phenomenological approach to test whether Pacini type scaling is still
applicable. Our approach has been to try and restrict the effects of
geometry by taking the peak luminosity as a scaling parameter rather
than the total luminosity. In this regard we are removing the duty
cycle term from PS87. Furthermore it is our opinion that to first
order the peak emission represents the local power density along the
observer's line of site.
               
\section{The Phenomenology of Magnetospheric Emission}

The three brightest pulsars (Crab, Vela and PSR 0540-69) are also
amongst the youngest. Table 1 shows the basic parameters for these
objects. However all the pulsars have very different pusle shapes
resulting in a very different ratio between the integrated flux and
the peak flux. Table 1 also shows this peak emission (taken as the
emission at the top of the largest peak). Their distances imply that
the thermal emission should be low (in all cases $<$ 1\% of the
observed emission).
       
Of all the optical pulsars PSR 0633+17 is perhaps the most
controversial. Early observations (Halpern \& Tytler (1988), and
Bignani et al (1988)) indicated that Geminga was an $\approx$ 25.5
m$_V$ object. Subsequent observations including HST photometry
appeared to support a thermal origin for the optical photons, albeit
requiring an arbitary assumption of cyclotron resonance feature in the
optical (Mignani et al, 1998).  The optical observations of Shearer et
al (1998) combined with spectroscopic observations (Martin et al,
1998) contradict this view. Figures 1 and 2 show how this
misunderstanding could have arisen. Figure 1, based upon data from
Bignani et al (1998) shows the integrated photometry. It would be
possible to fit a black body curve through this, but only with the
{\it a posteriori} fitting of a cyclotron resonance feature at about
5500 \AA. Figure 2 however shows the same point plotted on top of the
Martin et al spectra, we have also included the pulsed B point. This
combined data set indicates a flat spectrum consistent with
magnetosheric emission, without the requirement for such an {\it ad
hoc} feature. It was on the basis of these results that Golden \&
Shearer (1999) were able to give an upper limit of R$_\infty$ of about
10km.

\begin{figure}
\plottwo{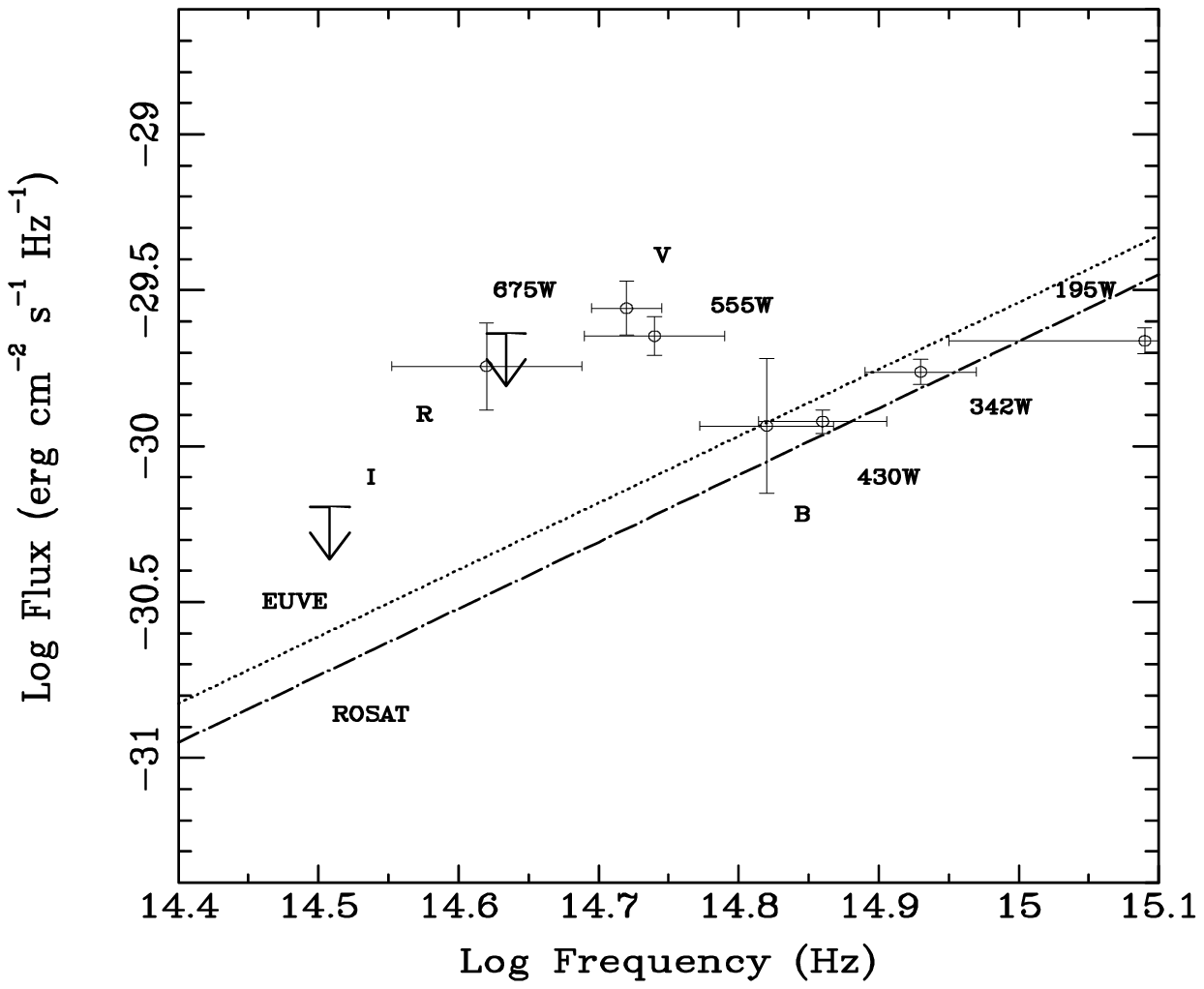}{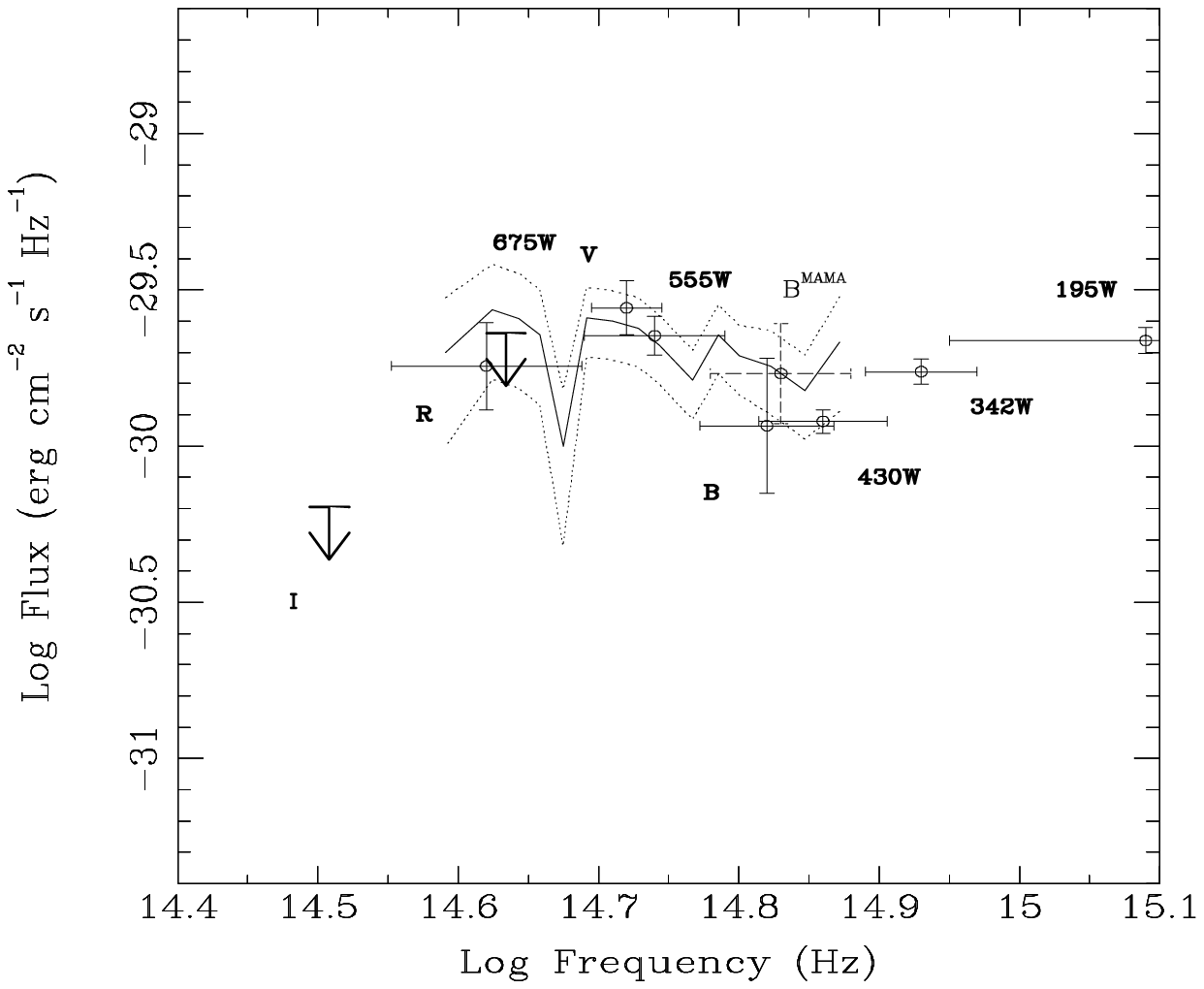}
\caption{\scriptsize{Spectra and Photometry of Geminga. The first
plot shows the integrated photometry and the thermal fit to the ROSAT
X-Ray data. The second illustrates the actual spectrum and the
agreement between it and the integrated photometry. B is the
pulsed flux from  Shearer et al (1998).\label{fits}} }
\end{figure}
               
With PSR 0656+14 there is a discrepancy between the radio distance
based upon the dispersion measure and the best fits to te X-Ray
data. From radio dispersion measure a distance of $ 760 \pm 190 pc $ can be
derived at odds with the X-ray distance of $250-280 pc$ from $N_H$
galactic models. Clearly more observations are needed to
determine a parallax.

Figure 2 shows the relationship between the peak luminosity with the
outer magnetic field. A regression of the form $Peak Luminoisty =
a*B^{b}$ was determined for the peak luminosity this lead to a
relationship of the form Peak Luminosity $ \propto B^{2.86 \pm 0.12}$
significant at the 99.5\% level. From PS87 we would expect a
relationship of the form Peak Luminosity $\propto B^{\approx 4}$ for
acceptable values of the energy spectrum exponent of the emitting
electrons - in reasonable agreement with our derived
relationship. Whilst informative it still goes no further than
previous attempts to understand the phenomena of optical emission. The
flattening of the peak luminosity relationship for the older, slower
pulsars is consistent with their having a steeper energy spectrum than
the younger pulsars. However we can state that from both polarisation
studies (Smith(1988);Romani \& Yadigaroglu (1995)) and from this work
we expect that optical emission zone should be sited towards the outer
magnetosphere. Timing studies of the size of the Crab pulse plateau
indicates a restricted emission volume ($\approx$ 45 kms in lateral
extent) (Golden et al (2000)). This third point, if consistent with
the first two, probably points to emission coming from a geometrically
defined cusp along our line of sight. Finally, there is no evidence of
optical thermal emission from these 5 pulsing optical neutron stars
(Martin et al (1998); this work).

\begin{figure}
\plotfiddle{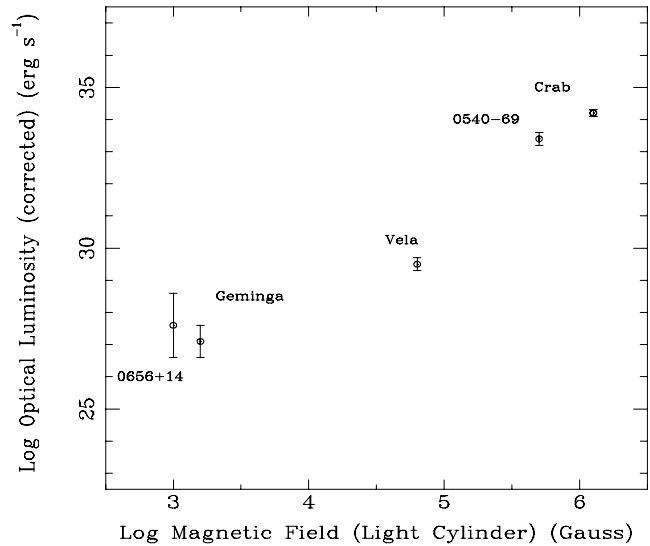}{6cm}{0}{70}{70}{-200}{-180}
\caption{\scriptsize{Outer field strength versus peak luminosity
\label{fits}}}
\end{figure}
             
\section{Conclusion}

Over the next few years (with the advent of larger telescopes and more
sensitive detectors, see for example Perryman (1999) \& Romani (1999))
we can confidently expect the number of optical detections of isolated
neutron stars to increase. In this region of the spectrum any potential
thermal component can be separated from the strongly pulsed magnetospheric
emission, allowing for reliable estimates of the neutron star radius to
be measured with consequent implications for equation of state
models. One word of caution however - our studies (see Golden et al
(1999) and this conference) indicate that the optical emission (at
least from the Crab pulsar) also exhibits an unpulsed component.

\begin{table}
\begin{scriptsize}

\caption{\scriptsize{Main Characteristics of Optical Pulsars: B$_S$ \& B$_{LC}$ the canonical surface and transverse magnetic field at the light cylinder respectively; Opt. Lum and Peak Lum. refer to the optical luminosity at the indicated distance in the B band}}
\begin{tabular}{lccccccc}

Name & D      & P  & $\dot{P}$  &  $B_S$ & $B_{LC}$  & Opt. Lum & Peak Lumin.\\
     &  (kpc) & (ms)   &  $10^{-14}$ s/s & log(G) & log (G)   & $\mu$Crab & $\mu$Crab \\

Crab        & 2       & 33  & 40  & 12.6 & 6.1  & $10^6$ & $10^6$ \\
Vela        & 0.5     & 89  & 11  & 12.5 & 4.8  & 27 & 21 \\
PSR0545-69  & 49      & 50  & 40  & 12.7 & 5.7  & $1.1~10^6$ & $1.4~10^5$ \\
PSR0656+14  & 0.76(?) & 385 & 1.2 & 12.7 & 3.0  & 1.8 & 0.3 \\
PSR0633+17     & 0.16    & 237 & 1.2 & 12.2 & 3.2  & 0.3 & 0.1 \\

\end{tabular}
\end{scriptsize}

\end{table}

\end{document}